\documentclass[aps,prb,amsmath,amssymb,superscriptaddress,twocolumn,longbibliography]{revtex4-2}
\usepackage{bbm}
\usepackage{graphicx}
\usepackage{dcolumn}
\usepackage{bm}
\usepackage{subfigure}
\usepackage{amsmath}
\usepackage{stmaryrd}
\usepackage{times}
\usepackage{graphicx}
\usepackage{dcolumn}
\usepackage{color}
\usepackage{makecell}
\usepackage{tabularx,multirow}
\usepackage{appendix}
\usepackage{amssymb}
\usepackage{pifont}
\usepackage[colorlinks,linkcolor=blue,citecolor=blue,urlcolor=black]{hyperref}
\begin{document}
\title{Highly anisotropic nonrelativistic charge-to-spin conversion in altermagnets}
\author{Mingbo Dou}
\affiliation{School of Physics, Harbin Institute of Technology, Harbin 150001, China}
\author{Xu Chen}
\affiliation{School of Physics, Harbin Institute of Technology, Harbin 150001, China}
\author{Zhiguo Liu}
\affiliation{School of Physics, Harbin Institute of Technology, Harbin 150001, China}
\author{Yu Sui}
\affiliation{School of Physics, Harbin Institute of Technology, Harbin 150001, China}
\author{Xianjie Wang}
\affiliation{School of Physics, Harbin Institute of Technology, Harbin 150001, China}
\affiliation{Heilongjiang Provincial Key Laboratory of Advanced Quantum Functional Materials and Sensor Devices, Harbin 150001, China}
\author{L. L. Tao}
\email{Contact author: lltao@hit.edu.cn}
\affiliation{School of Physics, Harbin Institute of Technology, Harbin 150001, China}
\affiliation{Heilongjiang Provincial Key Laboratory of Advanced Quantum Functional Materials and Sensor Devices, Harbin 150001, China}
\date{\today}
\begin{abstract}
The charge-to-spin conversion provides an efficient way to manipulate the magnetization by electrical means. In this work, we report on a study on the anisotropic nonrelativistic charge-to-spin conversion response to the current direction in altermagnets. We prove that the spin conductivity dictating the charge-to-spin conversion is equivalent to the spin-polarized conductivity difference in the nonrelativistic limit. Based on the general spin-group analysis, we derive analytical expressions for the anisotropic conversion ratio and identify its maximum value. We then exemplify those phenomena in representative altermagnets based on the density functional theory calculations. The highly anisotropic charge-to-spin conversion efficiency, varying from zero to several tens of percent, was demonstrated. Our work shines more light on the exploration of the nonrelativistic generation of spin currents in altermagnets.
\end{abstract}
\maketitle
\section{Introduction}
The generation of spin currents through the charge-to-spin conversion by electrical means is of vital importance in spintronics\cite{rmd1213,na509,npj27,nrm258}. Spin (polarized) currents can exert spin torques including spin-transfer torque (STT)\cite{jmmmL1,prb9353,prl3149,nm372} and spin-orbit torque (SOT)\cite{prb212405,prb094422,prb094424,nature189,science555,prl096602,rmp035004,apr011305} that may lead to the current-induced magnetization switching crucial for lower-power spintronic devices\cite{apl120502,pms100761,apr041316,apr041306,npj8,prl076801}. It is known that the spin-polarized current responsible for the STT is odd under the time reversal $\mathcal{T}$ while the spin current responsible for the SOT is even under $\mathcal{T}$. Despite significant advances in STT and SOT based spintronic devices, the inherent deficiencies have limited the further development of spin-torque devices such as nonvolatile magnetic random-access memories\cite{apl120502,pms100761}. For example, the spin current induced by the spin Hall effect\cite{rmd1213} or the Rashba effect\cite{jpcmR179,prb085438} is largely based on the strength of spin-orbit coupling (SOC), which is rather small as compared to the magnetic exchange coupling. Moreover, the SOC induced spin current typically reveals the in-plane spin polarization, which is not favorable to switch the perpendicular magnetization\cite{apl120502,pms100761,npj8}.

Recently, the spin-splitter torque (SST)\cite{prl127701} based on the spin-splitter effect was theoretically proposed in altermagnets\cite{prx031042,prx040501,afm2409327,jcas2257,nrm2025,nature837} endowed with the zero net magnetization and the sizable nonrelativistic spin splitting of the order of $1.0$ eV\cite{jpsj123702,prb014422,nc2846}. As distinct from the SOT, the spin current responsible for the SST is induced by the nonrelativistic spin splitting and is odd under $\mathcal{T}$\cite{prl127701}. Moreover, the spin polarization parallel to the Néel vector is highly controllable promising for the perpendicular magnetization switching\cite{npj8}. A highly charge-to-spin conversion ratio of $\sim28\%$ was theoretically demonstrated in the altermagnet RuO$_2$\cite{prl127701} and the SST with high efficiency was later observed in the RuO$_2$ films\cite{prl197202,prl137201,as2400967,nc5646,nc1309}. It was found that the direction of spin current strongly depends on the crystal orientation while the spin polarization is parallel to the Néel vector\cite{prl197202}. Moreover, the field-free switching of the perpendicular magnetization of the Co layer via the SST exerted by the spin current of the RuO$_2$ layer was achieved at room temperature\cite{prl137201}.

On the other hand, the charge-to-spin conversion is expected to reveal spatial anisotropy for different current directions, as evidenced by the anisotropic spin Hall effect in nonmagnetic hcp metals\cite{prl246602} and anisotropic charge-to-spin conversion in the topological semimetal SrIrO$_3$\cite{prbL220409}. For altermagnets, the highly anisotropic Fermi surface due to significant spin-split bands is expected to give rise to the anisotropic charge-to-spin conversion for the current along different crystal orientations. This is partially supported by the crystal axis dependence of the SST, which was observed in the RuO$_2$/Py bilayer structure with different oriented crystal planes. An efficient SST to drive the magnetization switching was observed in the RuO$_2$(100)/Py structure while the SST is absent in the RuO$_2$(110)/Py structure\cite{prl197202,prl137201}. However, the systematic theoretical study on the anisotropic spin current for altermagnets remains largely unexplored. In particular, the spatial anisotropy and analytical formulas for the nonrelativistic charge-to-spin conversion remains to be studied. It is the purpose of this work to explore the anisotropic charge-to-spin conversion and identify the crystal orientation sustaining the maximum conversion efficiency.

The rest of the paper is organized as follows. In Sec. \ref{sec2}, we present the computational method and details for the density functional theory (DFT) calculations. Section \ref{sec3} presents the theoretical formalism for the spin current and charge-to-spin conversion ratio calculations. In Sec. \ref{sec4}, we discuss the anisotropic charge-to-spin conversion based on the general symmetry analysis. In Sec. \ref{sec5}, we present the DFT results for representative altermagnets. Section \ref{sec6} is devoted to the anisotropic SST and spin injection due to the anisotropic spin current. Finally, Sec. \ref{sec7} is reserved for further discussion and conclusion.

\section{Computational method and details\label{sec2}}
\begin{figure}
\includegraphics[width=0.45\textwidth]{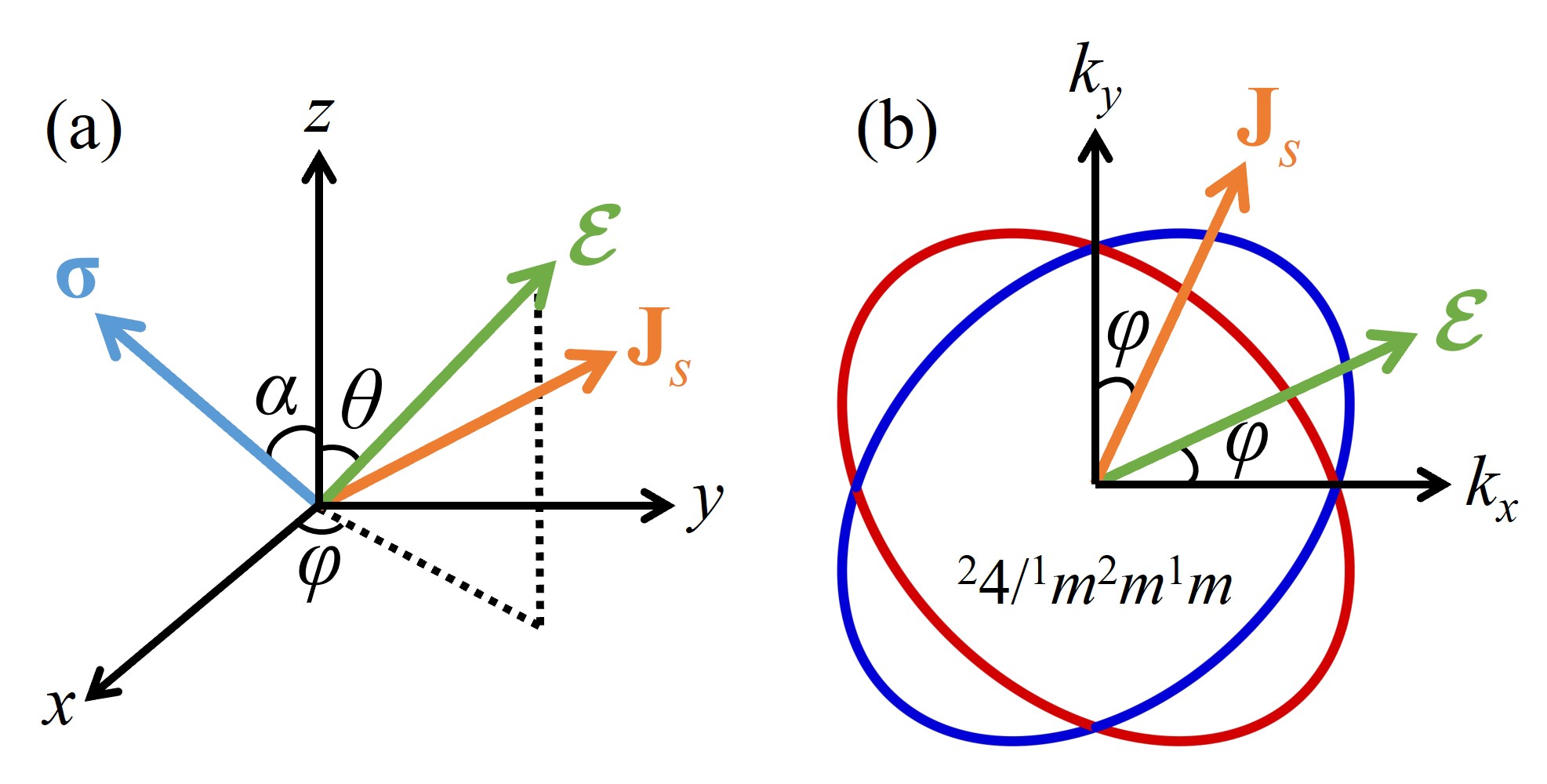}%
\caption{\label{f-1} (a) Sketches of the orientations of an electric field $\bm{\mathcal{E}}$, a spin current $\mathbf{J}_s$ and the spin polarization $\bm{\sigma}$ in Cartesian coordinates. $\theta$ and $\varphi$ indicate the polar angle and the azimuthal angle of $\bm{\mathcal{E}}$, respectively. $\alpha$ indicates the angle between the $i$-axis (here $i=z$) and $\bm{\sigma}$ (spin quantization axis). (b) Schematic illustration of the orientations of $\bm{\mathcal{E}}$ and $\mathbf{J}_s$ for the SPG $^24/^1m^2m^1m$. Red and blue contour lines indicate the spin-split Fermi contours. Note that the $(x, y, z)$ axis is concordant with the $(k_x, k_y, k_z)$ axis for the $d$-wave altermagnet.}
\end{figure}

The DFT calculations were performed using the plane-wave ultrasoft pseudopotential method\cite{prb7892} as implemented in the QUANTUM ESPRESSO\cite{jpcm395502,jpcm465901,jcp154105}. An energy cutoff of $50$ Ry for the plane-wave expansion and generalized gradient approximation (GGA)\cite{prl3865} for the exchange and correlation functional were adopted throughout. For the monoclinic CuF$_2$, we used the experimental lattice parameters\cite{jap1167} $a=3.30$, $b=4.57$ and $c=5.366$ {\AA} ($\beta=121.15^\text{o}$) and the $k$-point grid of $12\times10\times8$. A Hubbard-$U$ correction of $U_{eff}=4.0$ eV on the Cu-$d$ orbital was adopted to capture the Mott insulator\cite{ssc1703}. For the orthorhombic FeSb$_2$, we used the experimental lattice constants\cite{pnase2108924118} $a=5.834$, $b=6.53$ and $c=3.193$ {\AA} and the $k$-point grid of $8\times8\times16$. For the tetragonal K$_2$Ru$_8$O$_{16}$, we used the experimental lattice parameters\cite{jlcm323,prb195101} with lattice constants $a=9.866$, $c=3.131$ {\AA} and the $k$-point grid of $4\times4\times12$. For the tetragonal RuO$_2$, we used the experimental lattice constants\cite{prl077201} $a=4.492$, $c=3.106$ {\AA} and the $k$-point grid of $8\times8\times12$. A Hubbard-$U$ correction of $U_{eff}=3.0$ eV on the Ru-$4d$ orbital of K$_2$Ru$_8$O$_{16}$ and RuO$_2$ was adopted to capture the electron correlation\cite{prb195101}. For the orthorhombic MnPd$_2$\cite{jpcs212}, we used the lattice constant $a=8.123$, $b=5.481$ and $c=4.082$ {\AA} and the $k$-point grid of $5\times8\times10$. For the tetragonal OsO$_2$\cite{prm034407}, we used the lattice constants $a=4.492$, $c=3.106$ {\AA} and the $k$-point grid of $8\times8\times11$. A Hubbard-$U$ correction of $U_{eff}=3.0$ eV on the Os-$5d$ orbital was adopted to capture the electron correlation. The atomic positions for those altermagnets are listed in Supplemental Material, Table SI.

The spin conductivity was calculated using the Wannier-Linear-Response code\cite{prl187204,wtl} while the spin-resolved electrical conductivity was calculated by using the Boltzmann transport theory under the relaxation time approximation, as implemented in the BoltzWann module\cite{BoltzWann} of the Wannier90 code\cite{jpcm165902}. The tight-binding Hamiltonian was constructed from the maximally localized Wannier function basis set\cite{prb12847,rmp1419}. The $k$-point grids of $500\times350\times300$, $300\times300\times500$, $60\times60\times200$, $350\times350\times500$, $200\times320\times400$ and $280\times280\times400$ for evaluating the spin and electrical conductivities were used for CuF$_2$, FeSb$_2$, K$_2$Ru$_8$O$_{16}$, RuO$_2$, MnPd$_2$ and OsO$_2$, respectively.

\section{Theoretical formalism\label{sec3}}
\begin{figure*}
\includegraphics[width=0.9\textwidth]{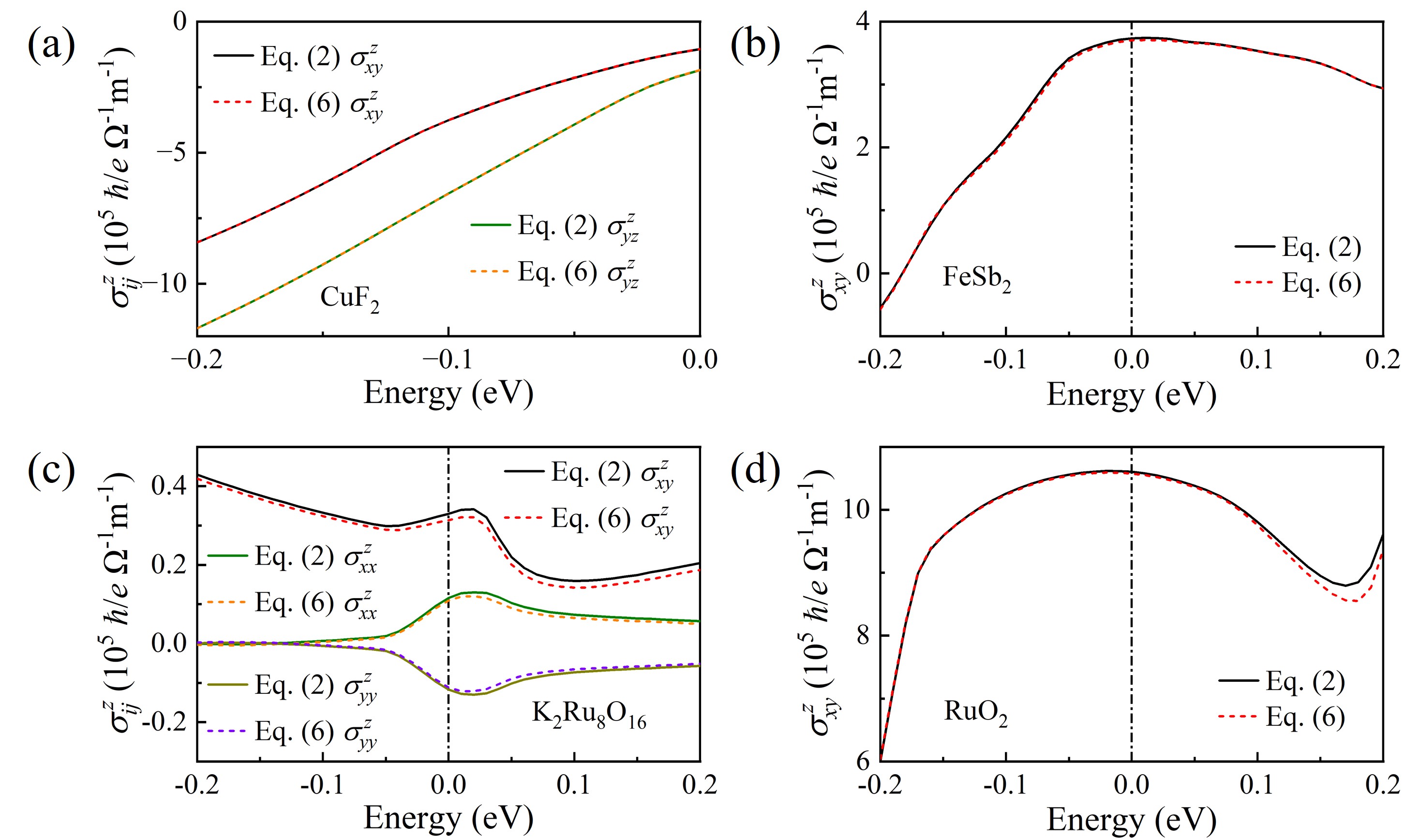}%
\caption{\label{f-2} $\mathcal{T}$-odd spin conductivities without SOC as a function of energy for the monoclinic CuF$_2$ (a), the orthorhombic FeSb$_2$ (b), the tetragonal K$_2$Ru$_8$O$_{16}$ (c) and the tetragonal RuO$_2$ (d). Solid lines were obtained using Eq. (\ref{eq-2}) with $\Gamma=10$ meV while dashed lines were obtained using Eq. (\ref{eq-6}) with $\tau=\hbar/(2\Gamma)\approx33$ fs (corresponding to $\Gamma=10$ meV) and a temperature of $50$ K in the Fermi distribution function. The Fermi energy (valence band maximum for CuF$_2$) has been aligned to zero.}
\end{figure*}
\begin{table*}
\caption{\label{table1} Symmetry dictated nonrelativistic spin conductivity components $\sigma_{jk}^i$ ($i, j, k=x, y, z$) and $\hat{\mathbf{n}}$-dependent spin-current direction $\hat{\mathbf{J}}_s$ for different SPGs sustaining the $d$-wave altermagnet.}
\begin{ruledtabular}
\begin{tabular}{cccccc}
SPG & $\sigma_{jk}^i$ & $\hat{\mathbf{J}}_s$ ($\hat{\mathbf{n}}||\hat{\mathbf{x}}$) & $\hat{\mathbf{J}}_s$ ($\hat{\mathbf{n}}||\hat{\mathbf{y}}$) & $\hat{\mathbf{J}}_s$ ($\hat{\mathbf{n}}||\hat{\mathbf{z}}$) & Candidate\\
\hline
$^22/^2m$ & $\sigma_{xy}^i$, $\sigma_{yz}^i$ & $-\text{sgn}(\bar{\sigma}_{xy}^\uparrow)\hat{\mathbf{y}}$ & $-\frac{\bar{\sigma}_{xy}^\uparrow\hat{\mathbf{x}}+\bar{\sigma}_{yz}^\uparrow\hat{\mathbf{z}}}{\sqrt{(\bar{\sigma}_{xy}^\uparrow)^2+(\bar{\sigma}_{yz}^\uparrow)^2}}$ & $-\text{sgn}(\bar{\sigma}_{yz}^\uparrow)\hat{\mathbf{y}}$ & CuF$_2$\\
$^2m^2m^1m$ & $\sigma_{xy}^i$ & $-\text{sgn}(\bar{\sigma}_{xy}^\uparrow)\hat{\mathbf{y}}$ & $-\text{sgn}(\bar{\sigma}_{xy}^\uparrow)\hat{\mathbf{x}}$ & - & FeSb$_2$, MnPd$_2$\\
$^24/^{1}m$ & $\sigma_{xx}^i$, $\sigma_{yy}^i$, $\sigma_{xy}^i$ & $-\frac{(\bar{\sigma}_{xx}^\uparrow-\bar{\sigma}_{xx}^\downarrow)\hat{\mathbf{x}}+2\bar{\sigma}_{xy}^\uparrow\hat{\mathbf{y}}}{\sqrt{(\bar{\sigma}_{xx}^\uparrow-\bar{\sigma}_{xx}^\downarrow)^2+4(\bar{\sigma}_{xy}^\uparrow)^2}}$ & $-\frac{2\bar{\sigma}_{xy}^\uparrow\hat{\mathbf{x}}-(\bar{\sigma}_{xx}^\uparrow-\bar{\sigma}_{xx}^\downarrow)\hat{\mathbf{y}}}{\sqrt{(\bar{\sigma}_{xx}^\uparrow-\bar{\sigma}_{xx}^\downarrow)^2+4(\bar{\sigma}_{xy}^\uparrow)^2}}$ & - & K$_2$Ru$_8$O$_{16}$\\
$^24/^1m^2m^1m$ & $\sigma_{xy}^i$ & $-\text{sgn}(\bar{\sigma}_{xy}^\uparrow)\hat{\mathbf{y}}$ & $-\text{sgn}(\bar{\sigma}_{xy}^\uparrow)\hat{\mathbf{x}}$ & - & RuO$_2$, OsO$_2$\\
\end{tabular}
\end{ruledtabular}
\end{table*}

Suppose that a conducting system is subject to an external electric field $\bm{\mathcal{E}}$. To first order in $\bm{\mathcal{E}}$, the spin current density $\mathbf{J}_s$ (second-rank tensor) takes the form\cite{prb024410}
\begin{equation}\label{eq-1}
  J_{sj}^i=\sigma_{jk}^i\mathcal{E}_k,
\end{equation}
where the spin conductivity $\sigma_{jk}^i$ (third-rank tensor) describes that a charge current along the $k$ direction induces a spin current along the $j$ direction with spin polarization along the $i$ direction. The indices $i, j, k=x, y, z$ denote Cartesian components and a summation over repeated indices is understood. Within the linear response theory using the Kubo formalism and the relaxation time $\tau$ approximation, two principal contributors to $\sigma_{jk}^i$ are the $\mathcal{T}$-odd contribution\cite{prb174423,na627,npj46}
\begin{equation}\label{eq-2}
    \sigma_{jk}^{i, \text{odd}} = -\frac{e\hbar}{\pi}\sum_{\mathbf{k},m,n}\frac{\text{Re}[\langle\psi_{n\mathbf{k}}|\mathcal{J}^i_j|\psi_{m\mathbf{k}}\rangle\langle\psi_{m\mathbf{k}}|v_k|\psi_{n\mathbf{k}}\rangle]\Gamma^2}{[(\epsilon_F-\epsilon_{n\mathbf{k}})^2+\Gamma^2][(\epsilon_F-\epsilon_{m\mathbf{k}})^2+\Gamma^2]},
\end{equation}
and the $\mathcal{T}$-even contribution (clean limit $\Gamma\rightarrow0$)
\begin{equation}\label{eq-3}
    \sigma_{jk}^{i, \text{even}} = -2e\hbar\sum_{\mathbf{k},m\neq n}\frac{\text{Im}[\langle\psi_{n\mathbf{k}}|\mathcal{J}^i_j|\psi_{m\mathbf{k}}\rangle\langle\psi_{m\mathbf{k}}|v_k|\psi_{n\mathbf{k}}\rangle]}{(\epsilon_{n\mathbf{k}}-\epsilon_{m\mathbf{k}})^2},
\end{equation}
where $\mathcal{J}^i_j=\hbar\{\sigma_i, v_j\}/4$ is the spin current operator given in terms of spin (velocity) operator $\sigma_i$ ($v_j$) and the reduced Planck's constant $\hbar$, $\epsilon_F$ the Fermi energy, $\Gamma$ the broadening parameter, and $\psi_{n\mathbf{k}}$ ($\epsilon_{n\mathbf{k}}$) the Bloch function (energy eigenvalue) of the $n$th band. $\tau$ and $\Gamma$ are related by $\tau=\hbar/(2\Gamma)$. Note that the formula Eq. (\ref{eq-2}) to calculate the $\mathcal{T}$-odd spin conductivity has been widely used by previous work\cite{prb174423,na627,npj46,prl127701,prx031042,prx040501}. In this work, we consider the nonrelativistic charge-to-spin conversion in altermagnets characterized by $\sigma_{jk}^{i, \text{odd}}$ due to that $\sigma_{jk}^{i, \text{even}}$ vanishes without SOC. Thus, the superscript ``odd'' in $\sigma_{jk}^{i, \text{odd}}$ is to be omitted in the following. In the clean limit of $\Gamma\rightarrow0$, by using the standard identity $\lim\limits_{\Gamma\to0}\frac{\Gamma}{\pi}\frac{1}{(\epsilon_F-\epsilon_{n\mathbf{k}})^2+\Gamma^2}=\delta(\epsilon_F-\epsilon_{n\mathbf{k}})$, Eq. (\ref{eq-2}) is reduced to\cite{prb174423}
\begin{equation}\label{eq-4}
    \sigma_{jk}^{i} = -e\tau\sum_{\mathbf{k}n}\langle\psi_{n\mathbf{k}}|\mathcal{J}^i_j|\psi_{n\mathbf{k}}\rangle\langle\psi_{n\mathbf{k}}|v_k|\psi_{n\mathbf{k}}\rangle\delta(\epsilon_F-\epsilon_{n\mathbf{k}}).
\end{equation}
It is to be noted that the spin polarization $i$ is dictated by the Néel vector orientation. For example, when the Néel vector is along the $z$ direction, we have $i=z$. With the Néel vector being in the $x-y$ plane, we have $i=x, y$, and so forth. For the collinear spin configuration without SOC, $|\psi_{n\mathbf{k}}\rangle$ can be denoted as $|\psi^s_{n\mathbf{k}}\rangle$ with $s=\uparrow, \downarrow$ being spin indices along the spin quantization axis. It is assumed that the angle between the $i$-axis and the spin quantization axis is $\alpha$ [see Fig. \ref{f-1}(a)]. Then, we have
\begin{equation}\label{eq-5}
\begin{aligned}
&\sum_{s}\langle\psi^s_{n\mathbf{k}}|\mathcal{J}^i_j|\psi^s_{n\mathbf{k}}\rangle=\langle\psi^\uparrow_{n\mathbf{k}}|\mathcal{J}^i_j|\psi^\uparrow_{n\mathbf{k}}\rangle+\langle\psi^\downarrow_{n\mathbf{k}}|\mathcal{J}^i_j|\psi^\downarrow_{n\mathbf{k}}\rangle\\
&=\frac{\hbar}{4}(\langle\psi^\uparrow_{n\mathbf{k}}|\{\sigma_i, v_j\}|\psi^\uparrow_{n\mathbf{k}}\rangle+\langle\psi^\downarrow_{n\mathbf{k}}|\{\sigma_i, v_j\}|\psi^\downarrow_{n\mathbf{k}}\rangle)\\
&=\frac{\hbar}{2}\cos\alpha(\langle\psi^\uparrow_{n\mathbf{k}}|v_j|\psi^\uparrow_{n\mathbf{k}}\rangle-\langle\psi^\downarrow_{n\mathbf{k}}|v_j|\psi^\downarrow_{n\mathbf{k}}\rangle)\\
&=\frac{\hbar}{2}\cos\alpha(v_{\mathbf{k}j}^{n\uparrow}-v_{\mathbf{k}j}^{n\downarrow}).
\end{aligned}
\end{equation}
This is substituted into Eq. (\ref{eq-4}), and the following result is obtained
\begin{equation}\label{eq-6}
    \sigma_{jk}^{i} = -\frac{\hbar}{2e}(\bar{\sigma}_{jk}^{\uparrow}-\bar{\sigma}_{jk}^{\downarrow})\cos\alpha,
\end{equation}
where the spin-resolved conductivity $\bar{\sigma}_{ij}^s$ reads\cite{callaway,prb155411,prb075422}
\begin{equation}\label{eq-7}
    \bar{\sigma}_{jk}^{s}(\epsilon_F) = e^2\tau\sum_{\mathbf{k}n}v_{\mathbf{k}j}^{ns}v_{\mathbf{k}k}^{ns}\delta(\epsilon_F-\epsilon_{n\mathbf{k}}).
\end{equation}

It follows from Eq. (\ref{eq-6}) that $\sigma_{jk}^i$ is proportional to the spin-polarized conductivity difference $\bar{\sigma}_{jk}^{\uparrow}-\bar{\sigma}_{jk}^{\downarrow}$ in the nonrelativistic limit. This can be confirmed by comparison of Eqs. (\ref{eq-2}) and (\ref{eq-6}) in the nonrelativistic limit. To examine this, we consider several prototypical altermagnets, that is, the monoclinic CuF$_2$, the orthorhombic FeSb$_2$, the tetragonal K$_2$Ru$_8$O$_{16}$ and the tetragonal RuO$_2$ and calculate $\sigma_{jk}^i$ using both Eqs. (\ref{eq-2}) and (\ref{eq-6}). Since the Néel vector is along the [001] axis ($z$ direction) for those altermagnets, we have $i=z$ and $\alpha=0$ in Eq. (\ref{eq-6}).

Figure \ref{f-2} shows $\sigma_{ij}^z$ ($i, j=x, y$) as a function of energy for the monoclinic CuF$_2$ [Fig. \ref{f-2}(a)], the orthorhombic FeSb$_2$ [Fig. \ref{f-2}(b)], the tetragonal K$_2$Ru$_8$O$_{16}$ [Fig. \ref{f-2}(c)] and the tetragonal RuO$_2$ [Fig. \ref{f-2}(d)]. It is seen that $\sigma_{ij}^z$ calculated from Eq. (\ref{eq-2}) is in substantial agreement with that calculated from Eq. (\ref{eq-6}). This confirms the validity of Eq. (\ref{eq-6}) for $\sigma_{jk}^i$ calculations in the nonrelativistic limit.

As evident from Eq. (\ref{eq-6}), in addition to $\alpha\neq\pi/2$, the nonzero $\sigma_{jk}^i$ requires that $\bar{\sigma}_{jk}^{\uparrow}\neq\bar{\sigma}_{jk}^{\downarrow}$, which can be determined in advance from symmetry arguments\cite{prb224423}. Given the fact that all the conductivity components for the $g$-wave and $i$-wave altermagnets are spin degenerate\cite{prb224423}, we focus on the $d$-wave altermagnet throughout this work. For an applied electric field $\bm{\mathcal{E}}$ along the direction $\hat{\mathbf{n}}=(\sin\theta\cos\varphi, \sin\theta\sin\varphi, \cos\theta)$ [$\theta$ for polar angle and $\varphi$ for azimuthal angle, see Fig. \ref{f-1}(a)], on combining Eqs. (\ref{eq-1}) and (\ref{eq-6}), we have
\begin{widetext}
\begin{equation}\label{eq-8}
\left(
  \begin{array}{c}
    J_{sx}^i \\
    J_{sy}^i \\
    J_{sz}^i \\
  \end{array}
\right)=-\frac{\hbar\cos\alpha}{2e}\left(
                  \begin{array}{c}
                    (\bar{\sigma}_{xx}^{\uparrow}-\bar{\sigma}_{xx}^{\downarrow})\sin\theta\cos\varphi+(\bar{\sigma}_{xy}^{\uparrow}-\bar{\sigma}_{xy}^{\downarrow})\sin\theta\sin\varphi+(\bar{\sigma}_{xz}^{\uparrow}-\bar{\sigma}_{xz}^{\downarrow})\cos\theta \\
                    (\bar{\sigma}_{xy}^{\uparrow}-\bar{\sigma}_{xy}^{\downarrow})\sin\theta\cos\varphi+(\bar{\sigma}_{yy}^{\uparrow}-\bar{\sigma}_{yy}^{\downarrow})\sin\theta\sin\varphi+(\bar{\sigma}_{yz}^{\uparrow}-\bar{\sigma}_{yz}^{\downarrow})\cos\theta \\
                    (\bar{\sigma}_{xz}^{\uparrow}-\bar{\sigma}_{xz}^{\downarrow})\sin\theta\cos\varphi+(\bar{\sigma}_{yz}^{\uparrow}-\bar{\sigma}_{yz}^{\downarrow})\sin\theta\sin\varphi+(\bar{\sigma}_{zz}^{\uparrow}-\bar{\sigma}_{zz}^{\downarrow})\cos\theta \\
                  \end{array}
                \right)\mathcal{E}=\bm{\sigma}(\theta, \varphi)\mathcal{E},
\end{equation}
\end{widetext}
where $\bm{\sigma}(\theta, \varphi)$ is a three-by-one column vector. From Eq. (\ref{eq-8}), one can specify the spin-current direction $\hat{\mathbf{J}}_s$ and the anisotropic spin conductivity $\bm{\sigma}(\theta, \varphi)$. A schematic illustration of $\bm{\mathcal{E}}$ induced $\mathbf{J}_s$ is shown in Fig. \ref{f-1}(b). In addition, the conductivity along the current direction $\hat{\mathbf{n}}$ is given by\cite{prb224423}
\begin{equation}\label{eq-9}
\begin{aligned}
    &\bar{\sigma}(\theta, \varphi)=\bar{\sigma}_{xx}\sin^2\theta\cos^2\varphi+\bar{\sigma}_{yy}\sin^2\theta\sin^2\varphi+\bar{\sigma}_{zz}\cos^2\theta\\
    &+\bar{\sigma}_{xy}\sin^2\theta\sin(2\varphi)+\bar{\sigma}_{yz}\sin(2\theta)\sin\varphi+\bar{\sigma}_{xz}\sin(2\theta)\cos\varphi,
\end{aligned}
\end{equation}
where we have used the fact that $\bar{\sigma}_{ij}=\bar{\sigma}_{ij}^{\uparrow}+\bar{\sigma}_{ij}^{\downarrow}$ for the collinear spin case. Then, on combining Eqs. (\ref{eq-8}) and (\ref{eq-9}), we introduce the anisotropic charge-to-spin conversion ratio $\Theta(\theta, \varphi)$ defined as
\begin{equation}\label{eq-10}
    \Theta(\theta, \varphi)=\frac{e}{\hbar}\frac{|\bm{\sigma}(\theta, \varphi)|}{\bar{\sigma}(\theta, \varphi)}.
\end{equation}
Without loss of generality, $\alpha=0$ in Eq. (\ref{eq-8}) will be used in the following.

\section{Group-theoretical analysis\label{sec4}}
\begin{table*}
\caption{\label{table2} Symmetry dictated anisotropic charge-to-spin conversion $\Theta(\theta, \varphi)$ and maximum conversion ratio $\Theta_{max}$ in the nonrelativistic case for different SPGs.}
\begin{ruledtabular}
\begin{tabular}{cccc}
SPG & $\Theta(\theta, \varphi)$ & $\Theta_{max}$ & Condition\\
\hline
$^22/^2m$ & $\frac{\sqrt{(\bar{\sigma}_{xy}^\uparrow\sin\theta)^2+(\bar{\sigma}_{yz}^\uparrow\cos\theta)^2+(\bar{\sigma}_{yz}^\uparrow\sin\theta\sin\varphi)^2+\bar{\sigma}_{xy}^\uparrow\bar{\sigma}_{yz}^\uparrow\sin(2\theta)\cos\varphi}}
{\bar{\sigma}_{xx}\sin^2\theta\cos^2\varphi+\bar{\sigma}_{yy}\sin^2\theta\sin^2\varphi+\bar{\sigma}_{zz}\cos^2\theta+\bar{\sigma}_{xz}\sin(2\theta)\cos\varphi}$ & - & -\\
\multirow{4}{*}{$^2m^2m^1m$} & \multirow{4}{*}{$\frac{|\bar{\sigma}_{xy}^\uparrow|\sin\theta}{\bar{\sigma}_{xx}\sin^2\theta\cos^2\varphi+\bar{\sigma}_{yy}\sin^2\theta\sin^2\varphi+\bar{\sigma}_{zz}\cos^2\theta}$} & $\frac{|\bar{\sigma}_{xy}^\uparrow|}{\bar{\sigma}_{xx}}$ & $\bar{\sigma}_{xx}\leq\bar{\sigma}_{yy}\leq2\bar{\sigma}_{zz}, \bar{\sigma}_{xx}\leq2\bar{\sigma}_{zz}\leq\bar{\sigma}_{yy}$\\
& & $\frac{|\bar{\sigma}_{xy}^\uparrow|}{\bar{\sigma}_{yy}}$ & $\bar{\sigma}_{yy}\leq\bar{\sigma}_{xx}\leq2\bar{\sigma}_{zz}, \bar{\sigma}_{yy}\leq2\bar{\sigma}_{zz}\leq\bar{\sigma}_{xx}$\\
& & $\frac{|\bar{\sigma}_{xy}^\uparrow|}{2\bar{\sigma}_{zz}}\sqrt{\frac{\bar{\sigma}_{zz}}{\bar{\sigma}_{xx}-\bar{\sigma}_{zz}}}$ & $\bar{\sigma}_{yy}>\bar{\sigma}_{xx}>2\bar{\sigma}_{zz}$\\
& & $\frac{|\bar{\sigma}_{xy}^\uparrow|}{2\bar{\sigma}_{zz}}\sqrt{\frac{\bar{\sigma}_{zz}}{\bar{\sigma}_{yy}-\bar{\sigma}_{zz}}}$ & $\bar{\sigma}_{xx}>\bar{\sigma}_{yy}>2\bar{\sigma}_{zz}$\\
\multirow{2}{*}{$^24/^{1}m$} & \multirow{2}{*}{$\frac{\sqrt{(\bar{\sigma}_{xx}^\uparrow-\bar{\sigma}_{xx}^\downarrow)^2+4(\bar{\sigma}_{xy}^\uparrow)^2}\sin\theta}{2(\bar{\sigma}_{xx}\sin^2\theta+\bar{\sigma}_{zz}\cos^2\theta)}$} & $\frac{\sqrt{(\bar{\sigma}_{xx}^\uparrow-\bar{\sigma}_{xx}^\downarrow)^2+4(\bar{\sigma}_{xy}^\uparrow)^2}}{2\bar{\sigma}_{xx}}$ & $\bar{\sigma}_{xx}\leq2\bar{\sigma}_{zz}$\\
& & $\sqrt{\frac{(\bar{\sigma}_{xx}^\uparrow-\bar{\sigma}_{xx}^\downarrow)^2+4(\bar{\sigma}_{xy}^\uparrow)^2}{16\bar{\sigma}_{zz}(\bar{\sigma}_{xx}-\bar{\sigma}_{zz})}}$ & $\bar{\sigma}_{xx}>2\bar{\sigma}_{zz}$ \\ 
\multirow{2}{*}{$^24/^1m^2m^1m$} & \multirow{2}{*}{$\frac{|\bar{\sigma}_{xy}^\uparrow|\sin\theta}{\bar{\sigma}_{xx}\sin^2\theta+\bar{\sigma}_{zz}\cos^2\theta}$} & $\frac{|\bar{\sigma}_{xy}^\uparrow|}{\bar{\sigma}_{xx}}$ & $\bar{\sigma}_{xx}\leq2\bar{\sigma}_{zz}$\\
& & $\frac{|\bar{\sigma}_{xy}^\uparrow|}{2\bar{\sigma}_{zz}}\sqrt{\frac{\bar{\sigma}_{zz}}{\bar{\sigma}_{xx}-\bar{\sigma}_{zz}}}$ & $\bar{\sigma}_{xx}>2\bar{\sigma}_{zz}$\\
\end{tabular}
\end{ruledtabular}
\end{table*}

We first analyze the anisotropic charge-to-spin conversion by use of the general spin-group analysis. It has been established that the altermagnet can be described by the third type of spin point groups (SPGs)\cite{prx031042,prx040501}. As mentioned above, only the $d$-wave altermagnet sustains the spin-polarized conductivity and the corresponding SPGs are $^22/^2m$, $^2m^2m^1m$, $^24/^{1}m$, and $^24/^1m^2m^1m$\cite{prx031042,prx040501}, as listed in Table \ref{table1}. For the SPG $^22/^2m$, the spin-polarized conductivity components are $\bar{\sigma}_{xy}$ and $\bar{\sigma}_{yz}$\cite{prb224423}. From Eq. (\ref{eq-8}), the spin-current direction $\hat{\mathbf{J}}_s$ can be obtained as
\begin{widetext}
\begin{equation}\label{eq-11}
    \hat{\mathbf{J}}_s=-\frac{\bar{\sigma}_{xy}^\uparrow\sin\theta\sin\varphi\hat{\mathbf{x}}+(\bar{\sigma}_{xy}^\uparrow\sin\theta\cos\varphi+\bar{\sigma}_{yz}^\uparrow\cos\theta)\hat{\mathbf{y}}+\bar{\sigma}_{yz}^\uparrow\sin\theta\sin\varphi\hat{\mathbf{z}}}
{\sqrt{(\bar{\sigma}_{xy}^\uparrow\sin\theta)^2+(\bar{\sigma}_{yz}^\uparrow\cos\theta)^2+(\bar{\sigma}_{yz}^\uparrow\sin\theta\sin\varphi)^2+\bar{\sigma}_{xy}^\uparrow\bar{\sigma}_{yz}^\uparrow\sin(2\theta)\cos\varphi}},
\end{equation}
and the anisotropic conversion ratio $\Theta$ can be obtained from Eq. (\ref{eq-10}) as
\begin{equation}\label{eq-12}
    \Theta=\frac{\sqrt{(\bar{\sigma}_{xy}^\uparrow\sin\theta)^2+(\bar{\sigma}_{yz}^\uparrow\cos\theta)^2+(\bar{\sigma}_{yz}^\uparrow\sin\theta\sin\varphi)^2+\bar{\sigma}_{xy}^\uparrow\bar{\sigma}_{yz}^\uparrow\sin(2\theta)\cos\varphi}}
{\bar{\sigma}_{xx}\sin^2\theta\cos^2\varphi+\bar{\sigma}_{yy}\sin^2\theta\sin^2\varphi+\bar{\sigma}_{zz}\cos^2\theta+\bar{\sigma}_{xz}\sin(2\theta)\cos\varphi}.
\end{equation}
\end{widetext}
$\hat{\mathbf{J}}_s$ for different $\hat{\mathbf{n}}$ can be immediately inferred from Eq. (\ref{eq-12}). In Table \ref{table1}, we summarize $\hat{\mathbf{J}}_s$ when $\hat{\mathbf{n}}$ is along one of the three principal axes. It is easily seen that, for $\hat{\mathbf{n}}||\hat{\mathbf{x}}$ and $\hat{\mathbf{n}}||\hat{\mathbf{z}}$, $\hat{\mathbf{J}}_s$ is along the $y$ direction as expected from the nonzero components $\sigma_{xy}^i$ and $\sigma_{yz}^i$. For $\hat{\mathbf{n}}||\hat{\mathbf{y}}$, $\hat{\mathbf{J}}_s$ is in the $x-z$ plane. On the other hand, the extrema of $\Theta$ can be determined in principle from that $\partial\Theta/\partial\theta=0$ and $\partial\Theta/\partial\varphi=0$. However, the result is too complicated to evaluate analytically. This has to be analyzed by numerical methods.

For the SPG $^2m^2m^1m$, only $\bar{\sigma}_{xy}$ is spin polarized\cite{prb224423}. Similarly, we have
\begin{equation}\label{eq-13}
    \hat{\mathbf{J}}_s=-\text{sgn}(\bar{\sigma}_{xy}^\uparrow\sin\theta)(\sin\varphi\hat{\mathbf{x}}+\cos\varphi\hat{\mathbf{y}}),
\end{equation}
and
\begin{equation}\label{eq-14}
    \Theta=\frac{|\bar{\sigma}_{xy}^\uparrow|\sin\theta}{\bar{\sigma}_{xx}\sin^2\theta\cos^2\varphi+\bar{\sigma}_{yy}\sin^2\theta\sin^2\varphi+\bar{\sigma}_{zz}\cos^2\theta}.
\end{equation}
From Eq. (\ref{eq-13}), we see that $\hat{\mathbf{J}}_s$ is always in the $x-y$ plane and perpendicular to $\hat{\mathbf{n}}$ for $\hat{\mathbf{n}}||\hat{\mathbf{x}}$ and $\hat{\mathbf{n}}||\hat{\mathbf{y}}$. From Eq. (\ref{eq-14}), the extrema of $\Theta$ can be examined analytically from $\partial\Theta/\partial\theta=0$ and $\partial\Theta/\partial\varphi=0$, which yields the following $12$ stationary points $\hat{\mathbf{n}}=(\theta, \varphi)$
\begin{equation}\label{eq-15}
\begin{aligned}
     &\hat{\mathbf{n}}_1=(\frac{\pi}{2}, 0), \hat{\mathbf{n}}_2=(\frac{\pi}{2}, \pi), \hat{\mathbf{n}}_3=(\frac{\pi}{2}, \frac{\pi}{2}), \hat{\mathbf{n}}_4=(\frac{\pi}{2}, \frac{3\pi}{2}),\\
     &\hat{\mathbf{n}}_5=(\theta_1, 0), \hat{\mathbf{n}}_6=(\pi-\theta_1, 0), \hat{\mathbf{n}}_7=(\theta_1, \pi),\\
     &\hat{\mathbf{n}}_8=(\pi-\theta_1, \pi), \hat{\mathbf{n}}_9=(\theta_2, \frac{\pi}{2}), \hat{\mathbf{n}}_{10}=(\pi-\theta_2, \frac{\pi}{2}),\\
     &\hat{\mathbf{n}}_{11}=(\theta_2, \frac{3\pi}{2}), \hat{\mathbf{n}}_{12}=(\pi-\theta_2, \frac{3\pi}{2}),\\
\end{aligned}
\end{equation}
where $\theta_{1, 2}$ are defined as $\theta_1\equiv\arcsin\sqrt{\bar{\sigma}_{zz}/(\bar{\sigma}_{xx}-\bar{\sigma}_{zz})}$, $\theta_2\equiv\arcsin\sqrt{\bar{\sigma}_{zz}/(\bar{\sigma}_{yy}-\bar{\sigma}_{zz})}$. 
We substitute Eq. (\ref{eq-15}) into Eq. (\ref{eq-14}) and obtain the maximum value $\Theta_{max}$ by comparing $\Theta$'s at different stationary points
\begin{equation}\label{eq-16}
\begin{aligned}
    &\Theta_{max}^{\hat{\mathbf{n}}_{1,2}}=\frac{|\bar{\sigma}_{xy}^\uparrow|}{\bar{\sigma}_{xx}}, \bar{\sigma}_{xx}\leq\bar{\sigma}_{yy}\leq2\bar{\sigma}_{zz}~\text{or}~\bar{\sigma}_{xx}\leq2\bar{\sigma}_{zz}\leq\bar{\sigma}_{yy},\\
    &\Theta_{max}^{\hat{\mathbf{n}}_{3,4}}=\frac{|\bar{\sigma}_{xy}^\uparrow|}{\bar{\sigma}_{yy}}, \bar{\sigma}_{yy}\leq\bar{\sigma}_{xx}\leq2\bar{\sigma}_{zz}~\text{or}~\bar{\sigma}_{yy}\leq2\bar{\sigma}_{zz}\leq\bar{\sigma}_{xx},\\
    &\Theta_{max}^{\hat{\mathbf{n}}_{5,6,7,8}}=\frac{|\bar{\sigma}_{xy}^\uparrow|}{2\bar{\sigma}_{zz}}\sqrt{\frac{\bar{\sigma}_{zz}}{\bar{\sigma}_{xx}-\bar{\sigma}_{zz}}}, \bar{\sigma}_{yy}>\bar{\sigma}_{xx}>2\bar{\sigma}_{zz},\\
    &\Theta_{max}^{\hat{\mathbf{n}}_{9,10,11,12}}=\frac{|\bar{\sigma}_{xy}^\uparrow|}{2\bar{\sigma}_{zz}}\sqrt{\frac{\bar{\sigma}_{zz}}{\bar{\sigma}_{yy}-\bar{\sigma}_{zz}}}, \bar{\sigma}_{xx}>\bar{\sigma}_{yy}>2\bar{\sigma}_{zz}.\\
\end{aligned}
\end{equation}

For the SPG $^24/^{1}m$, the conductivity components $\bar{\sigma}_{xx}$, $\bar{\sigma}_{yy}$ and $\bar{\sigma}_{xy}$ are spin polarized\cite{prb224423}. A similar argument yields
\begin{widetext}
\begin{equation}\label{eq-17}
    \hat{\mathbf{J}}_s=-\text{sgn}(\sin\theta)\frac{[(\bar{\sigma}_{xx}^\uparrow-\bar{\sigma}_{xx}^\downarrow)\cos\varphi+2\bar{\sigma}_{xy}^\uparrow\sin\varphi]\hat{\mathbf{x}}+[2\bar{\sigma}_{xy}^\uparrow\cos\varphi-(\bar{\sigma}_{xx}^\uparrow-\bar{\sigma}_{xx}^\downarrow)\sin\varphi]\hat{\mathbf{y}}}
{\sqrt{(\bar{\sigma}_{xx}^\uparrow-\bar{\sigma}_{xx}^\downarrow)^2+4(\bar{\sigma}_{xy}^\uparrow)^2}},
\end{equation}
\end{widetext}
and
\begin{equation}\label{eq-18}
    \Theta=\frac{\sqrt{(\bar{\sigma}_{xx}^\uparrow-\bar{\sigma}_{xx}^\downarrow)^2+4(\bar{\sigma}_{xy}^\uparrow)^2}\sin\theta}{2(\bar{\sigma}_{xx}\sin^2\theta+\bar{\sigma}_{zz}\cos^2\theta)}.
\end{equation}
Equation (\ref{eq-17}) suggests that the spin current is always in the $x-y$ plane. As distinct from the SPGs $^22/^2m$ and $^2m^2m^1m$, $\Theta$ is only dependent on $\theta$, as seen from Eq. (\ref{eq-18}). We then find the following stationary points from $\partial\Theta/\partial\theta=0$
\begin{equation}\label{eq-19}
    \theta_{1,2,3}=\frac{\pi}{2}, \arcsin\sqrt{\frac{\bar{\sigma}_{zz}}{\bar{\sigma}_{xx}-\bar{\sigma}_{zz}}}, \pi-\arcsin\sqrt{\frac{\bar{\sigma}_{zz}}{\bar{\sigma}_{xx}-\bar{\sigma}_{zz}}}.
\end{equation}
We use Eqs. (\ref{eq-18}) and (\ref{eq-19}) and find, with the same substitutions,
\begin{equation}\label{eq-20}
\begin{aligned}
    &\Theta_{max}^{\theta_1}=\frac{\sqrt{(\bar{\sigma}_{xx}^\uparrow-\bar{\sigma}_{xx}^\downarrow)^2+4(\bar{\sigma}_{xy}^\uparrow)^2}}{2\bar{\sigma}_{xx}}, \bar{\sigma}_{xx}\leq2\bar{\sigma}_{zz},\\
    &\Theta_{max}^{\theta_{2,3}}=\sqrt{\frac{(\bar{\sigma}_{xx}^\uparrow-\bar{\sigma}_{xx}^\downarrow)^2+4(\bar{\sigma}_{xy}^\uparrow)^2}{16\bar{\sigma}_{zz}(\bar{\sigma}_{xx}-\bar{\sigma}_{zz})}}, \bar{\sigma}_{xx}>2\bar{\sigma}_{zz}.
\end{aligned}
\end{equation}

For the SPG $^24/^1m^2m^1m$, only $\bar{\sigma}_{xy}$ is spin polarized\cite{prb224423}. We substitute in Eqs. (\ref{eq-8}) and (\ref{eq-10}) and find
\begin{equation}\label{eq-21}
    \hat{\mathbf{J}}_s=-\text{sgn}(\bar{\sigma}_{xy}^\uparrow\sin\theta)(\sin\varphi\hat{\mathbf{x}}+\cos\varphi\hat{\mathbf{y}}),
\end{equation}
and
\begin{equation}\label{eq-22}
    \Theta=\frac{|\bar{\sigma}_{xy}^\uparrow|\sin\theta}{\bar{\sigma}_{xx}\sin^2\theta+\bar{\sigma}_{zz}\cos^2\theta},
\end{equation}
which is $\varphi$ independent. We see that Eq. (\ref{eq-21}) is the same with Eq. (\ref{eq-13}) due to the same nonzero spin conductivity component $\sigma_{xy}^i$ between the SPGs $^2m^2m^1m$ and $^24/^1m^2m^1m$. Imposing the condition $\partial\Theta/\partial\theta=0$ yields the following stationary points
\begin{equation}\label{eq-23}
    \theta_{1,2,3}=\frac{\pi}{2}, \arcsin\sqrt{\frac{\bar{\sigma}_{zz}}{\bar{\sigma}_{xx}-\bar{\sigma}_{zz}}}, \pi-\arcsin\sqrt{\frac{\bar{\sigma}_{zz}}{\bar{\sigma}_{xx}-\bar{\sigma}_{zz}}}.
\end{equation}
In this case, $\Theta_{max}$ can be obtained as
\begin{equation}\label{eq-24}
\begin{aligned}
    &\Theta_{max}^{\theta_1}=\frac{|\bar{\sigma}_{xy}^\uparrow|}{\bar{\sigma}_{xx}}, \bar{\sigma}_{xx}\leq2\bar{\sigma}_{zz},\\
    &\Theta_{max}^{\theta_{2,3}}=\frac{|\bar{\sigma}_{xy}^\uparrow|}{2\bar{\sigma}_{zz}}\sqrt{\frac{\bar{\sigma}_{zz}}{\bar{\sigma}_{xx}-\bar{\sigma}_{zz}}}, \bar{\sigma}_{xx}>2\bar{\sigma}_{zz}.
\end{aligned}
\end{equation}

\begin{table*}
\caption{\label{table3} Symmetry dictated nonrelativistic spin conductivity components $\sigma_{jk}^i$ ($i=x, y, z$ and $j, k=x, y$) and $\hat{\mathbf{n}}$-dependent spin-current direction $\hat{\mathbf{J}}_s$ for different spin layer groups sustaining the $d$-wave two-dimensional altermagnet. $\Theta(\varphi)$ and $\Theta_{max}$ indicate the anisotropic and the maximum charge-to-spin conversion ratios, respectively.}
\begin{ruledtabular}
\begin{tabular}{cccccc}
Spin layer group & $\sigma_{jk}^i$ ($i=x, y, z$) & $\hat{\mathbf{J}}_s$ ($\hat{\mathbf{n}}||\hat{\mathbf{x}}$) & $\hat{\mathbf{J}}_s$ ($\hat{\mathbf{n}}||\hat{\mathbf{y}}$) & $\Theta(\varphi)$ & $\Theta_{max}$\\
\hline
\multirow{2}{*}{$^22/^2m$, $^2m^2m^1m$} & \multirow{2}{*}{$\sigma_{xy}^i$} & \multirow{2}{*}{$-\text{sgn}(\bar{\sigma}_{xy}^\uparrow)\hat{\mathbf{y}}$} & \multirow{2}{*}{$-\text{sgn}(\bar{\sigma}_{xy}^\uparrow)\hat{\mathbf{x}}$} & \multirow{2}{*}{$\frac{|\bar{\sigma}_{xy}^\uparrow|}{\bar{\sigma}_{xx}\cos^2\varphi+\bar{\sigma}_{yy}\sin^2\varphi}$} & $\frac{|\bar{\sigma}_{xy}^\uparrow|}{\bar{\sigma}_{xx}}$, $\bar{\sigma}_{xx}\leq\bar{\sigma}_{yy}$\\
& & & & & $\frac{|\bar{\sigma}_{xy}^\uparrow|}{\bar{\sigma}_{yy}}$, $\bar{\sigma}_{xx}>\bar{\sigma}_{yy}$\\
$^24/^{1}m$ & $\sigma_{xx}^i$, $\sigma_{yy}^i$, $\sigma_{xy}^i$ & $-\frac{(\bar{\sigma}_{xx}^\uparrow-\bar{\sigma}_{xx}^\downarrow)\hat{\mathbf{x}}+2\bar{\sigma}_{xy}^\uparrow\hat{\mathbf{y}}}{\sqrt{(\bar{\sigma}_{xx}^\uparrow-\bar{\sigma}_{xx}^\downarrow)^2+4(\bar{\sigma}_{xy}^\uparrow)^2}}$ & $-\frac{2\bar{\sigma}_{xy}^\uparrow\hat{\mathbf{x}}-(\bar{\sigma}_{xx}^\uparrow-\bar{\sigma}_{xx}^\downarrow)\hat{\mathbf{y}}}{\sqrt{(\bar{\sigma}_{xx}^\uparrow-\bar{\sigma}_{xx}^\downarrow)^2+4(\bar{\sigma}_{xy}^\uparrow)^2}}$ & $\frac{\sqrt{(\bar{\sigma}_{xx}^\uparrow-\bar{\sigma}_{xx}^\downarrow)^2+4(\bar{\sigma}_{xy}^\uparrow)^2}}{2\bar{\sigma}_{xx}}$ & $\frac{\sqrt{(\bar{\sigma}_{xx}^\uparrow-\bar{\sigma}_{xx}^\downarrow)^2+4(\bar{\sigma}_{xy}^\uparrow)^2}}{2\bar{\sigma}_{xx}}$\\ 
$^24/^1m^2m^1m$ & $\sigma_{xy}^i$ & $-\text{sgn}(\bar{\sigma}_{xy}^\uparrow)\hat{\mathbf{y}}$ & $-\text{sgn}(\bar{\sigma}_{xy}^\uparrow)\hat{\mathbf{x}}$ & $\frac{|\bar{\sigma}_{xy}^\uparrow|}{\bar{\sigma}_{xx}}$ & $\frac{|\bar{\sigma}_{xy}^\uparrow|}{\bar{\sigma}_{xx}}$\\
\end{tabular}
\end{ruledtabular}
\end{table*}

In Table \ref{table2}, we summarize the above analytic expressions for the anisotropic charge-to-spin conversion $\Theta(\theta, \varphi)$ and the maximum conversion ratio $\Theta_{max}$ for different SPGs. It indicates that $\Theta(\theta, \varphi)$ is dictated by the SPG while $\Theta_{max}$ is material dependent, which is reflected by the comparison of the diagonal conductivity components $\bar{\sigma}_{ii}$. Thus, the optimal charge-to-spin conversion responsible for the STT can be achieved through engineering the orientation between the electrical current and the crystal axis.

Up to the present, our discussion of the symmetry dictated charge-to-spin conversion is focused on bulk altermagnets. The theory can also be extended in a straightforward way to describe the two-dimensional altermagnets\cite{apl182409,cs13853,jcas381,afme17921}, which are described by the spin layer group\cite{prb054406}. Namely, we have $\theta=\pi/2$ in charge-to-spin conversion expressions $\Theta(\theta, \varphi)$ and the conversion anisotropy is described by $\varphi$. In Table \ref{table3}, we summarize the spin-layer-group dictated nonrelativistic spin conductivity components,  the anisotropic $\Theta(\varphi)$ and the maximum $\Theta_{max}$ charge-to-spin conversion ratios. As distinct from bulk altermagnets summarized in Table \ref{table1}, the spin layer groups $^22/^2m$ and $^2m^2m^1m$ yield the same spin conductivity components and charge-to-spin conversion ratios. Another major difference is that $\Theta$ is $\varphi$ independent for the spin layer groups $^24/^{1}m$ and $^24/^1m^2m^1m$, which indicates that the charge-to-spin conversion is isotropic.

\section{DFT results\label{sec5}}

\begin{figure*}
\includegraphics[width=0.9\textwidth]{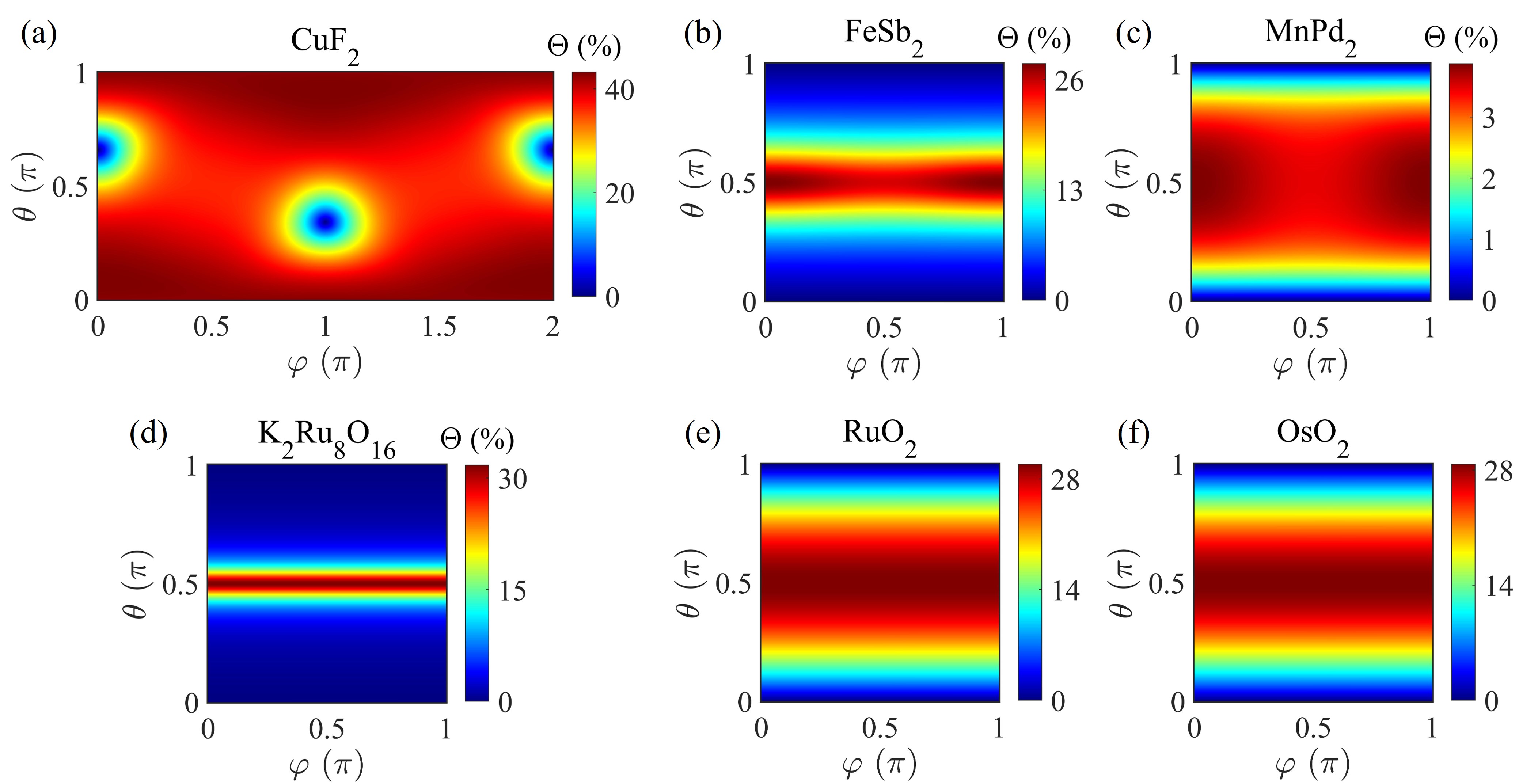}%
\caption{\label{f-3} Anisotropic charge-to-spin conversion ratio $\Theta$ at the Fermi energy as a function of $(\theta, \varphi)$ for CuF$_2$ (a), FeSb$_2$ (b), MnPd$_2$ (c), K$_2$Ru$_8$O$_{16}$ (d), RuO$_2$ (e) and OsO$_2$ (f). Note that, the Fermi energy for the semiconductor CuF$_2$ is at $50$ meV below the valence band maximum.}
\end{figure*}

\begin{figure*}
\includegraphics[width=0.9\textwidth]{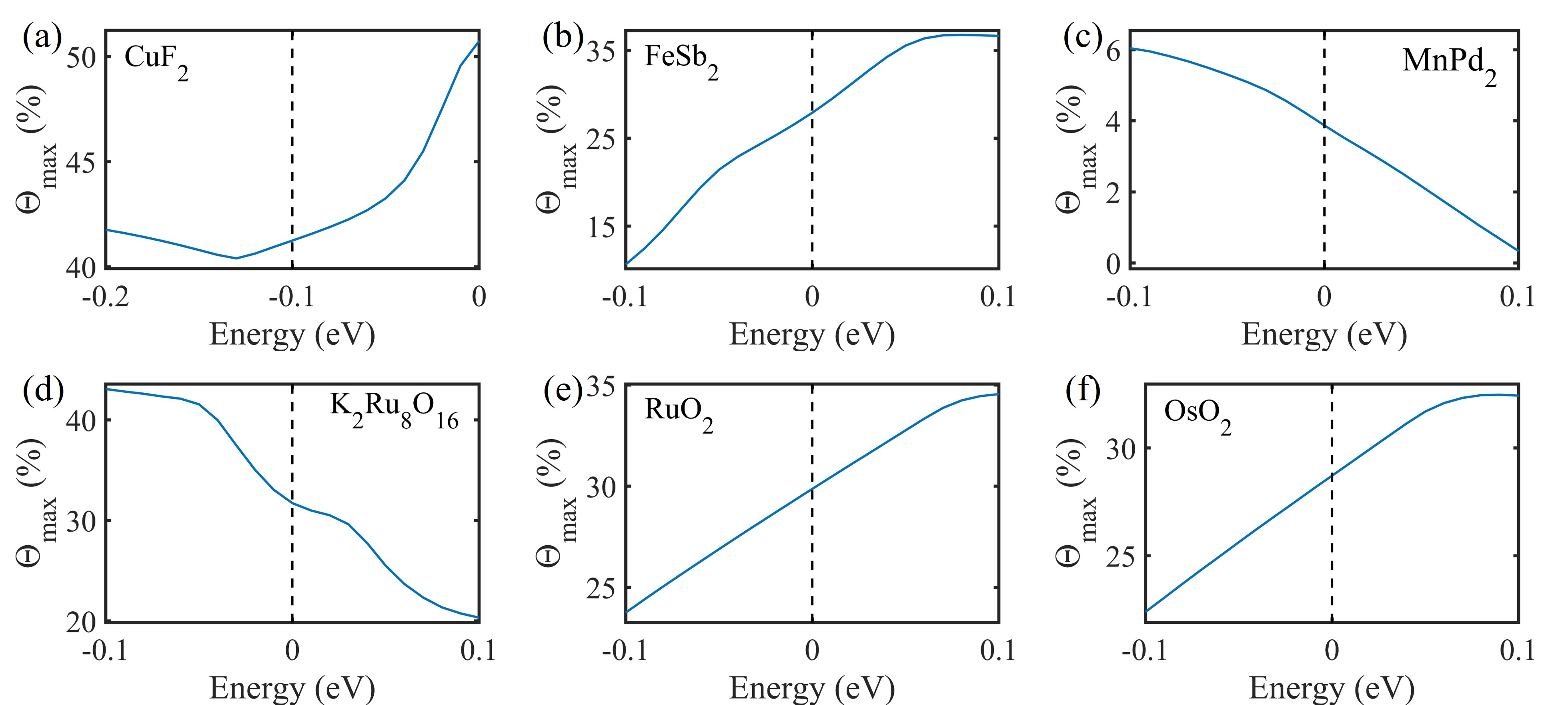}%
\caption{\label{f-4} Maximum charge-to-spin conversion ratio $\Theta_{max}$ as a function of energy for CuF$_2$ (a), FeSb$_2$ (b), MnPd$_2$ (c), K$_2$Ru$_8$O$_{16}$ (d), RuO$_2$ (e) and OsO$_2$ (f). The Fermi energy (valence band maximum for CuF$_2$) has been aligned to zero.}
\end{figure*}

Having demonstrated the anisotropic charge-to-spin conversion in $d$-wave altermagnets based on the general group-theoretical analysis, we now exemplify those phenomena in representative altermagnets based on the DFT calculations. We consider a handful of candidates, as listed in the last column of Table \ref{table1}. To be specific, we consider the monoclinic CuF$_2$ (SPG $^22/^2m$), the orthorhombic FeSb$_2$, MnPd$_2$ (SPG $^2m^2m^1m$), the tetragonal K$_2$Ru$_8$O$_{16}$ (SPG $^24/^{1}m$) and the tetragonal RuO$_2$, OsO$_2$ (SPG $^24/^1m^2m^1m$). Given the fact the CuF$_2$ is a wide-gap semiconductor while the other candidates are metallic, we assume that the Fermi energy is at the $50$ meV below the valence band maximum for CuF$_2$. It is also worth noting here that although the presence of magnetic order in the RuO$_2$ is highly controversial\cite{prl166702,prl176401}, we will, however, consider here the RuO$_2$ as an illustration of group-theoretical results.

Figure \ref{f-3} shows the anisotropic charge-to-spin conversion ratio $\Theta$ at the Fermi energy as a function of $(\theta, \varphi)$. It has to be stressed that Fig. \ref{f-3} was produced by using $\Theta (\theta, \varphi)$ formulas given by Eqs. (\ref{eq-12}), (\ref{eq-14}), (\ref{eq-18}) and (\ref{eq-22}), where the spin-polarized conductivity $\bar{\sigma}_{ij}^\uparrow$ was calculated under the relaxation time approximation. For the monoclinic CuF$_2$ shown in Fig. \ref{f-3}(a), it is indicated that $\Theta$ reveals large values of more than $40\%$ in a large area of the $\theta-\varphi$ plane except for three small regions with rather small values. This is to be expected since both the in-plane component $\bar{\sigma}_{xy}^\uparrow$ and the out-of-plane one $\bar{\sigma}_{yz}^\uparrow$ contribute to $\Theta$, as seen from Eq. (\ref{eq-12}). For the orthorhombic FeSb$_2$ shown in Fig. \ref{f-3}(b), $\Theta$ is significant around $\theta=\pi/2$, i.e. electric field along the in-plane direction. This is legitimate since only the in-plane spin conductivity component $\sigma^i_{xy}$ is symmetry allowed. In addition, $\Theta$ reaches the maximum value at $(\pi/2, 0)$ and $(\pi/2, \pi)$. This is due to the fact that $\bar{\sigma}_{xx}<\bar{\sigma}_{yy}<2\bar{\sigma}_{zz}$ at the Fermi energy. As a result, we have $\Theta_{max}=|\bar{\sigma}_{xy}^\uparrow|/\bar{\sigma}_{xx}\sim26.6\%$ at $(\pi/2, 0)$ and $(\pi/2, \pi)$ according to Eq. (\ref{eq-16}). A similar profile to FeSb$_2$ is observed for the orthorhombic MnPd$_2$, as shown in Fig. \ref{f-3}(c). However, the distribution of $\Theta (\theta, \varphi)$ is broadening and the value of $\Theta$ is significantly smaller than that of FeSb$_2$.

Figure \ref{f-3}(d) shows $\Theta (\theta, \varphi)$ for the tetragonal K$_2$Ru$_8$O$_{16}$. As distinct from CuF$_2$, FeSb$_2$ and MnPd$_2$, $\Theta$ is $\varphi$ independent as seen from Eq. (\ref{eq-18}). Since we have $\bar{\sigma}_{xx}<2\bar{\sigma}_{zz}$ at the Fermi energy, $\Theta$ reaches its maximum of more than $30\%$ at $\theta=\pi/2$ according to Eq. (\ref{eq-20}). A particularly noteworthy conversion ratio is centered on $\theta=\pi/2$ and $\Theta$ decays rapidly as $\theta$ is away from $\pi/2$. This can be understood as follows. Since the in-plane lattice constant of $9.866$ Å is significantly larger than the out-of-plane lattice constant of $3.131$ Å for K$_2$Ru$_8$O$_{16}$, the band structure reveals a less (strong) in-plane (out-of-plane) dispersion\cite{prb224423}, which yields $\bar{\sigma}_{zz}\approx30\bar{\sigma}_{xx}$. Thus, Eq. (\ref{eq-18}) can be simplified as $\Theta\sim\sin\theta/\cos^2\theta$, which reaches maximum at $\theta=\pi/2$ and decays quite quickly as $\theta$ deviates from $\pi/2$. Figure \ref{f-3}(e) shows $\Theta (\theta, \varphi)$ for the tetragonal RuO$_2$. It is seen that $\Theta$ is $\varphi$ independent similar to that for K$_2$Ru$_8$O$_{16}$. According to Eq. (\ref{eq-24}), since $\bar{\sigma}_{xx}<2\bar{\sigma}_{zz}$ at the Fermi energy, we have $\Theta_{max}=|\bar{\sigma}_{xy}^\uparrow|/\bar{\sigma}_{xx}\sim29.5\%$ at $\theta=\pi/2$. This value is quite similar to previous work\cite{prl127701,prb214419}. Another feature is that $\Theta (\theta)$ is rather broadening around $\theta=\pi/2$, contrary to that for K$_2$Ru$_8$O$_{16}$ [see Fig. \ref{f-3}(d)]. This illustrates the point that a large charge-to-spin conversion ratio can be achieved even if the charge current deviates greatly from the in-plane direction. Since the difference between the in-plane lattice constant of $4.492$ Å and the out-of-plane lattice constant of $3.106$ Å for RuO$_2$ is smaller than that for K$_2$Ru$_8$O$_{16}$\cite{prb224423}, we have $\bar{\sigma}_{xx}\approx\bar{\sigma}_{zz}$. Thus, Eq. (\ref{eq-22}) for RuO$_2$ is reduced to $\Theta\sim\sin\theta$, which reaches maximum at $\theta=\pi/2$ but decays slowly as $\theta$ deviates from $\pi/2$. A quite similar profile to RuO$_2$ is also observed for the tetragonal OsO$_2$, as shown in Fig. \ref{f-3}(f).

It is to be noted that the nonzero spin conductivity components can be intuitively understandable in terms of the anisotropic spin-split Fermi surface. As shown in Supplemental Material, Fig. S1, the spin-dependent Fermi surface is highly inequivalent along the diagonal direction, which yields the nonzero $\sigma_{xy}^i$ component for FeSb$_2$ and RuO$_2$.

Figure \ref{f-4} shows the maximum charge-to-spin conversion ratio $\Theta_{max}$ as a function of energy. It is indicated that there is no significant change in $\Theta_{max}$ near the Fermi energy. For the semiconductor CuF$_2$ shown in Fig. \ref{f-4}(a), $\Theta_{max}$ increases significantly when the energy approaches to the valence band maximum. $\Theta_{max}$ for FeSb$_2$ increases monotonously with increasing energy, as seen from Fig. \ref{f-4}(b). An opposite trend [see Fig. \ref{f-4}(c)] is observed for MnPd$_2$ although they have the same SPG. From Fig. \ref{f-4}(d), $\Theta_{max}$ for K$_2$Ru$_8$O$_{16}$ decreases monotonously with increasing energy. For RuO$_2$ and OsO$_2$, a quite similar energy dependency of $\Theta_{max}$ can be seen from Figs. \ref{f-4}(e) and \ref{f-4}(f).

\section{Anisotropic STT and spin injection\label{sec6}}

\begin{figure*}
\includegraphics[width=0.9\textwidth]{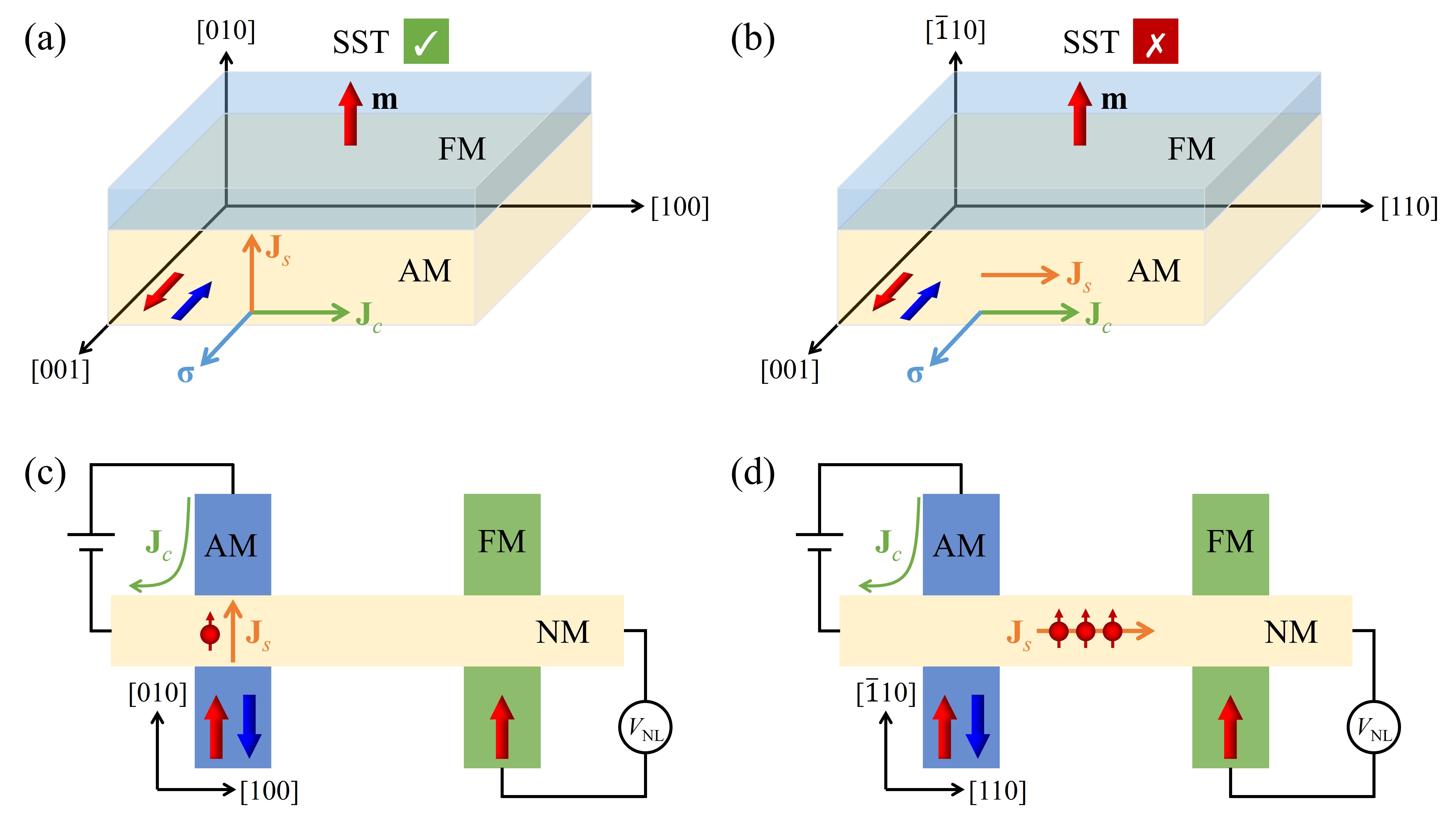}%
\caption{\label{f-5} (a, b) Sketches of the ferromagnet (FM)/altermagnet (AM) bilayer with different crystal orientations. $\mathbf{m}$ indicates the magnetization while $\mathbf{J}_c$ and $\mathbf{J}_s$ stand for the charge current and spin current, respectively. In configuration (a), $\mathbf{J}_s$ can generate an SST on $\mathbf{m}$ while the SST is absent in configuration (b). (c, d) Sketches of the lateral spin valve for the spin-injection detection with different crystal orientations. The spin-injection signal in configuration (d) is more pronounced than that in configuration (c) due to distinct nonrelativistic spin current directions.}
\end{figure*}

It is enlightening to examine the utility of the highly anisotropic charge-to-spin conversion in spintronic applications. Here we consider two promising applications, that is, the SST and the spin injection. We recall that the SST was proposed based on the spin-splitter effect in altermagnets, which can be described by the $\mathcal{T}$-odd spin conductivity\cite{prl127701}. Both the spin polarization direction and the spin current direction are highly controllable through controlling the Néel vector and the charge current directions\cite{prl197202,prl137201,as2400967,nc5646,nc1309}. Our predicted highly anisotropic charge-to-spin conversion can be directly reflected in the crystal axis dependence of the SST or the spin injection. To have a specific example, we consider the SPG $^2m^2m^1m$. As inspired by previous works\cite{prl127701,prl197202,prl137201}, from Eq. (\ref{eq-13}), the spin current is in the $x-y$ plane and the charge current direction $\hat{\mathbf{n}}$ sustaining $\Theta_{max}$ is determined by the comparison of $\bar{\sigma}_{xx}$, $\bar{\sigma}_{yy}$ and $2\bar{\sigma}_{zz}$ as given by Eq. (\ref{eq-16}). As shown in Fig. \ref{f-5}(a), the charge current $\mathbf{J}_c$ along the $[100]$ axis ($x$-direction) generates the nonrelativistic spin current along the $[010]$ axis ($y$-direction), which can exert an efficient spin torque to switch the magnetization of the adjacent ferromagnetic layer. In stark contrast, as shown in Fig. \ref{f-5}(b), the spin current is parallel with the charge current when the electric field is along the $[110]$ axis. In such a case, no spin current flows into the adjacent ferromagnetic layer and thus the SST is absent.

Another confirmation of the anisotropic charge-to-spin conversion is through examining the crystal axis dependence of the spin injection from the altermagnets into nonmagnetic metal\cite{arXiv:2512.17427}. Again, we consider the altermagnet with the SPG $^2m^2m^1m$ for the injection electrode. Figure \ref{f-5}(c) and (d) show the lateral spin valve structure for magnetotransport measurements. As seen from Fig. \ref{f-5}(c), the charge current along the $[100]$ axis yields a nonrelativistic spin current along the $[010]$ axis. Such nonequilibrium spin accumulation diffuses towards the ferromagnetic detection electrode, which may induce a non-local voltage $V_{NL}$. When the charge current is along the $[110]$ axis shown in Fig. \ref{f-5}(d), the spin current along the $[110]$ axis result in the spin accumulation observed via $V_{NL}$. More importantly, the spin signal observed in this geometry [Fig. \ref{f-5}(d)] is expected to be significantly larger than that for the former geometry [Fig. \ref{f-5}(c)] due to distinct spin current flowing directions.

\section{Discussion and summary\label{sec7}}
Note that the SOC is ignored since we focus on the nonrelativistic charge-to-spin conversion in this work. However, the effect of SOC on the conversion ratio would be negligible for the altermagnet without containing heavy elements. As summarized in Supplemental Material, Table SII, the difference between the maximum conversion ratio $\Theta_{max}$ without SOC and that with SOC for most altermagnets is negligible. An exception occurs for OsO$_2$ that the difference is slightly large as expected from the strong SOC of the heavy element Os. By considering the SOC, there are additional nonzero spin conductivity components in additional to those listed in Table \ref{table1}. In such cases, it will usually be convenient to dictate the spin conductivity by use of the magnetic point group\cite{prl127701,npj46}.

It is worth noting here that the nonrelativistic charge-to-spin conversion cannot be dictated by the spin polarization\cite{prb224423}, although the nonzero spin conductivity components are in line with spin-polarized conductivity components. This can be confirmed from the expression for the anisotropic charge-to-spin conversion ratio, which is in stark contrast to that for the anisotropic spin polarization\cite{prb224423}. For example, as seen from Eq. (\ref{eq-22}) and Fig. \ref{f-3}(e), the conversion ratio for RuO$_2$ is only $\theta$ dependent and reaches maximum at $\theta=\pi/2$. In stark contrast, the anisotropic spin polarization is both $\theta$ and $\varphi$ dependent and could be zero at $\theta=\pi/2$, as seen from Eq. (11) and Fig. 4(e) in Ref. \cite{prb224423}.

In summary, we have investigated the anisotropic nonrelativistic charge-to-spin conversion in altermagnets based on the group-theoretical analysis and DFT calculations. We show that the charge-to-spin conversion efficiency is highly anisotropic for the current along different crystal orientations. In particular, we derive the analytical expression for the anisotropic charge-to-spin conversion ratio and identify the direction sustaining the maximum conversion efficiency. It is also proved that the $\mathcal{T}$-odd spin conductivity is equivalent to the spin-polarized conductivity difference in the nonrelativistic limit, which significantly simplifies the nonrelativistic spin-conductivity expression. Our results are expected to pave the practical way to produce the high charge-to-spin conversion efficiency in altermagnets.

\begin{center}
{\bf SUPPLEMENTARY MATERIAL}
\end{center}
See the supplementary material for the atomic positions (Table SI), the comparision between the maximum conversion ratio without SOC and that with SOC (Table SII), and the anisotropic Fermi surfaces for FeSb$_2$ and RuO$_2$ (Fig. S1).

\begin{center}
{\bf ACKNOWLEDGMENTS}
\end{center}
This research was supported by the National Natural Science Foundation of China (Grant No. 12274102). The Fermi surfaces are produced using the FermiSurfer software\cite{FermiSurfer}.

\begin{center}
{\bf AUTHOR DECLARATIONS}
\end{center}

\begin{center}
{\bf Conflict of Interest}
\end{center}
The authors have no conflicts to disclose.

\begin{center}
{\bf DATA AVAILABILITY}
\end{center}
The data that support the findings of this study are available from the corresponding author upon reasonable request.

\end{document}